\documentclass[12pt]{article}

\usepackage{fancybox}

\usepackage{cite}
\usepackage{float}
\usepackage{amsfonts}
\usepackage{amsmath}
\usepackage{amsbsy}
\usepackage{graphicx}
\usepackage{amssymb}
\usepackage{amsthm}
\usepackage{bm}
\usepackage{epsfig}
\usepackage{latexsym}
\usepackage{pdflscape}
\usepackage{color}
\usepackage{here}
\usepackage{graphicx}
\numberwithin{equation}{section}

\allowdisplaybreaks

\DeclareMathOperator{\re}{Re}
\DeclareMathOperator{\im}{Im}
\DeclareMathOperator{\llangle}{\langle\!\langle\!}
\DeclareMathOperator{\rrangle}{\!\rangle\!\rangle}

\newcounter{aff}

\begin{document}

\begin{titlepage}
\begin{flushright}
{\footnotesize OCU-PHYS 437}
\end{flushright}
\bigskip
\begin{center}
{\LARGE\bf Orthosymplectic Chern-Simons Matrix Model\\[6pt]
and Chirality Projection}\\
\bigskip\bigskip
{\large 
Sanefumi Moriyama\footnote{\tt moriyama@sci.osaka-cu.ac.jp}
\quad and \quad
Takao Suyama\footnote{\tt suyama@sci.osaka-cu.ac.jp}
}\\
\bigskip
${}^*${\it Department of Physics, Graduate School of Science,
Osaka City University}\\
${}^\dagger${\it Osaka City University
Advanced Mathematical Institute (OCAMI)}\\
{\it 3-3-138, Sugimoto, Sumiyoshi, Osaka 558-8585, Japan}
\end{center}

\bigskip

\begin{abstract}

Recently it was found that the density matrix for a certain orthosymplectic Chern-Simons theory matches with that for the ABJM theory with the odd chiral projection.
We prove this fact for a general case with the inclusion of fractional branes.
We also identify the first few diagonal Gopakumar-Vafa invariants for the grand potential constructed from the chirally projected density matrix.

\centering\includegraphics[scale=0.75,bb=200 400 400 600]{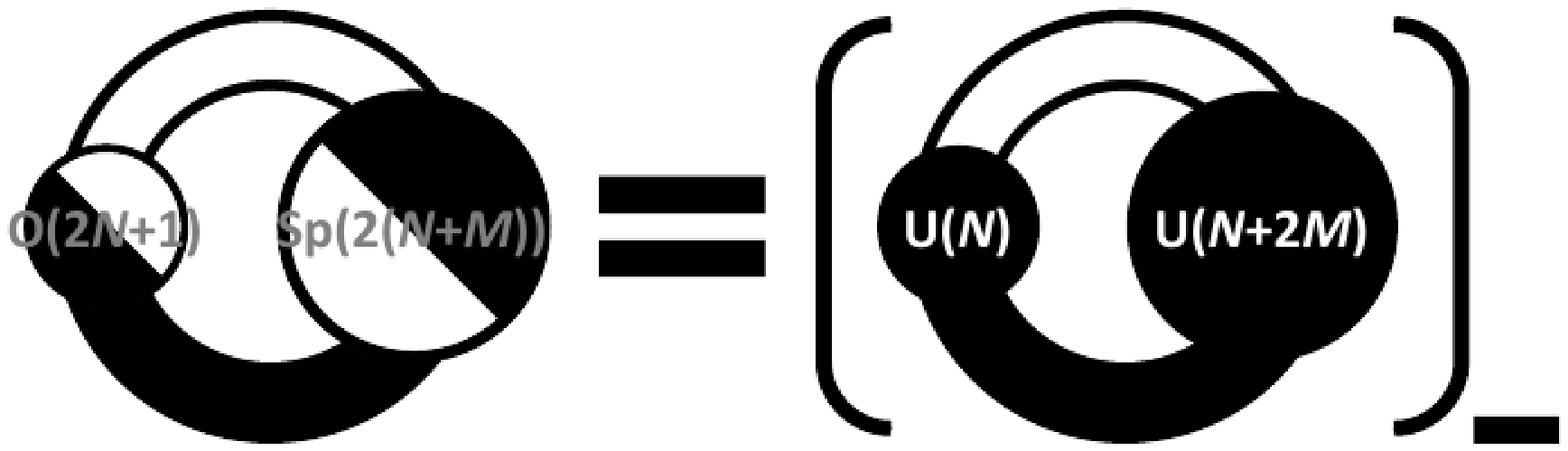}

\end{abstract}
\end{titlepage}

\tableofcontents

\section{Introduction}\label{intro}
The partition functions of three-dimensional Chern-Simons theories show various interesting aspects of M2-branes.
The would-volume theory of $N$ M2-branes on ${\mathbb R}^8/{\mathbb Z}_k$ is described by an ${\cal N}=6$ superconformal Chern-Simons theory called ABJM theory \cite{ABJM}, which has a gauge group U$(N)_k\times$U$(N)_{-k}$ (with the subscripts denoting the Chern-Simons levels) and two pairs of bifundamental matters connecting the two U$(N)$ factors.
Due to the progress in the supersymmetric localization \cite{KWY}, the partition function on a sphere is reduced to a matrix model with a finite-dimensional multiple integral.
One of the major developments is the full determination of the partition function of the ABJM theory in the large $N$ expansion, including the perturbative \cite{DMP1,DMP2,FHM,MP} and non-perturbative \cite{HMO2,CM,HMO3,HMMO} effects.
In the study, among others, it is interesting to find that the matrix model has several interpretations.
On one hand, it can be superficially regarded as the pure Chern-Simons matrix model with an unconventional super gauge group U$(N|N)$ \cite{DT}.
On the other hand, the matrix model can be regarded as the partition function of a Fermi gas system \cite{MP}
\begin{align}
Z^\text{ABJM}_k(N)=\frac{1}{N!}\sum_{\sigma\in S_N}(-1)^\sigma
\int\frac{d^N\mu}{(2\pi)^N}\prod_{i=1}^N
\langle\mu_i|\widehat\rho_{\text{U}(N|N)}|\mu_{\sigma(i)}\rangle,
\label{trace}
\end{align}
with a non-trivial density matrix
\begin{align}
\widehat\rho_{\text{U}(N|N)}
=\frac{1}{\sqrt{2\cosh\frac{\widehat q}{2}}}
\frac{1}{2\cosh\frac{\widehat p}{2}}
\frac{1}{\sqrt{2\cosh\frac{\widehat q}{2}}},
\label{density}
\end{align}
which is closely related to the quantum mechanical system associated to the local ${\mathbb P}^1\times{\mathbb P}^1$ geometry \cite{MPtop}.

It is then interesting to ask whether we can generalize the results to theories with a large number of supersymmetries.\footnote{
For other generalizations whose exact large $N$ expansion is known, see \cite{MN1,MN2,MN3} for the $(2,2)$ model and \cite{GHM1,OZ,Ha} for the local ${\mathbb P}^2$ model.}
One direction is the generalization to the matrix model with a superficial gauge group U$(N_1|N_2)$ \cite{HLLLP2,ABJ} where two factors of the bosonic subgroup have different ranks and the physical interpretation of the difference is the introduction of fractional M2-branes \cite{ABJ}.
In studying the partition function with the deformation \cite{AHS,Ho1,MM,HoOk,HNS}, there are two formulations.
The first one, called closed string formalism in \cite{PTEP}, changes the expression of the density matrix $\widehat\rho$ \eqref{density} while preserving the trace structure \eqref{trace}.
This formalism was first conjectured in \cite{AHS} and then proved in \cite{Ho1}.
Partially due to the lack of a proof of the formalism for a long time, in \cite{MM} another formalism, called open string formalism, was proposed.
This formalism, on the other hand, keeps the expression of the density matrix \eqref{density}, while modifying the trace structure \eqref{trace} with an extra determinant factor.

Another direction is the replacement of the unitary supergroup by the orthosymplectic supergroup \cite{HLLLP2,ABJ}, whose physical interpretation is the introduction of the orientifold plane in the type IIB description.
The study of the partition function was initiated in\footnote{
Some works which may be related to a similar physical setup are \cite{MN4,ADF,Ok1}.
}
\cite{MS1} by the case of OSp$(2N|2N)$ with equal sizes of bosonic submatrices from the expectation that the case without the fractional branes should play a fundamental role.
Among others it was found that the density matrix for this theory is closely related to $\bigl[\widehat\rho_{\text{U}(N|N)}\bigr]_+$, the density matrix for the ABJM theory with a projection to the even chirality.
Here the chirally projected density matrices
\begin{align}
\bigl[\widehat\rho_{\text{U}(N|N)}\bigr]_\pm
=\widehat\rho_{\text{U}(N|N)}\frac{1\pm\widehat R}{2},
\end{align}
were introduced in \cite{HMO1,MePu} with $\widehat R$ being the reflection operator changing the sign of the coordinate.
Then, it was found that when we double the quivers following the prescription in \cite{HoMo}, the partition function schematically reduces to the ABJM partition function.

Recently, there appeared an interesting paper \cite{Ho2}.
In \cite{Ho2}, it was observed that the OSp$(2N+1|2N)$ theory, still having equal ranks and hence no fractional branes \cite{ABJ}, seems to serve an equally fundamental role.
It was found that the density matrix for the OSp$(2N+1|2N)$ theory is exactly that of the ABJM theory with the projection to the odd chirality
\begin{align}
\widehat\rho_{\text{OSp}(2N+1|2N)}
=\bigl[\widehat\rho_{\text{U}(N|N)}\bigr]_-.
\label{oddproj}
\end{align}

\begin{figure}
\centering\includegraphics[scale=0.6,bb=200 350 400 500]{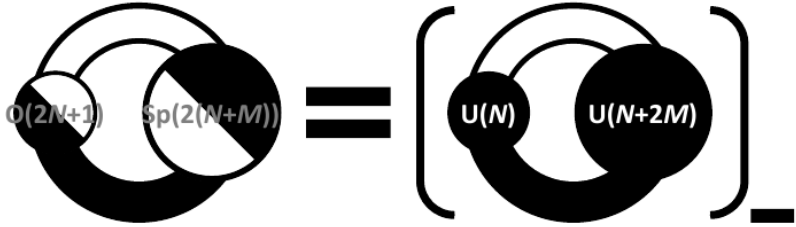}
\caption{A schematic relation between the density matrix for the orthosymplectic theory and that for the unitary theory.}
\label{oddpic}
\end{figure}

It is then interesting to ask whether and how this relation holds in the deformation into the case of different ranks.
The first part of this paper is devoted to answering this question.
We have found that, when we deform the theory into that with a superficial gauge group OSp$(2N+1|2(N+M))$ (or OSp$(2(N+M)+1|2N)$ which shares the same partition function), the density matrix is again exactly the odd projection of the density matrix for the theory with a superficial unitary gauge group U$(N|N+2M)$: 
\begin{align}
\widehat\rho_{\text{OSp}(2N+1|2(N+M))}
=\bigl[\widehat\rho_{\text{U}(N|N+2M)}\bigr]_-.
\label{oddprojM}
\end{align}
See figure \ref{oddpic} for a schematic picture explaining the relation.
We stress that the relation \eqref{oddprojM} gives a Fermi gas formalism for the OSp$(2N+1|2(N+M))$ theory, which enables the study of the grand potential and its relation to topological string theory.

Our manipulations start with an expression rather similar to the open string formalism \cite{MM}.
It is useful to keep the determinant factor coming from the open string formalism to see many cancellations in the expressions.
After performing a similarity transformation and an integration of delta functions, we can put the expression into the form of the closed string formalism and prove the relation \eqref{oddprojM}.
In both of the U$(N|N+2M)$ and OSp$(2N+1|2(N+M))$ theories there is a physical bound \cite{ABJ} stating that $0\le 2M\le k$.\footnote{
Note that the level in the orthosymplectic matrix model is $k$ instead of $2k$.
In other words, the number of D5-branes in the brane construction of \cite{ABJ} is $k$ in our convention.}
It is interesting to find that our relation between these two theories is consistent with the bound.
We stress that, although we are influenced by the work of \cite{Ho1}, 
it seems difficult to arrive at our proof of the relation \eqref{oddprojM} if we simply follow the change of variables in \cite{Ho1}.

Following the observation \eqref{oddproj}, in the second part, we turn to the study of the simplest $M=0$ case, the OSp$(2N+1|2N)$ theory, which is equivalent to the ABJM U$(N|N)$ theory with the odd chiral projection.
We study the exact values of the partition functions constructed from the chirally projected density matrices and read off the grand potentials $J_{\pm,k}(\mu)$ from the numerical fitting.
We find an interesting functional relation stating that the difference between $J_{+,k}(\mu)$ and $J_{-,k}(\mu)$ is extremely simple for integral $k$, with an explicit relation expressed in $k$ mod $8$ as in the case of the OSp$(2N|2N)$ theory \cite{MS1}.
We further turn to the worldsheet instanton effects and identify the diagonal Gopakumar-Vafa invariants.

This paper is organized as follows.
In section \ref{odd}, we present a proof for \eqref{oddprojM}.
After establishing this relation, we turn to the study of the grand potential in section \ref{functional}.
Finally we conclude with some discussions.
The appendix is devoted to a collection of several data which are needed for our claim in section \ref{functional}.

\noindent
{\bf Note Added}:
After this work was done and while we are preparing the draft, \cite{Ok2} appears on arXiv, which has some overlaps with our section \ref{functional} (especially \eqref{npdiff}).

\section{Orthosymplectic matrix model as odd projection}\label{odd}

In this section we shall prove that the density matrix for the orthosymplectic matrix model with the superficial gauge group OSp$(2N_1+1|2N_2)$ is equivalent to a chiral half of that for a matrix model with a suitable unitary super gauge group.

Let us start with the partition function of the orthosymplectic theory\footnote{
Compared with the standard normalization in the literature such as \cite{MS1}, the integral variables $\mu_i$ and $\nu_k$ are rescaled by $k$ from the beginning.}
\begin{align}
&Z_{k}(N_1,N_2)
=\int\frac{D^{N_1}\mu}{N_1!}\frac{D^{N_2}\nu}{N_2!}
\frac{V_\text{O}V_\text{Sp}}{H},
\end{align}
where the integration from the tree-level contribution is
\begin{align}
D\mu_i=\frac{d\mu_i}{4\pi k}e^{\frac{i}{4\pi k}\mu_i^2},\quad
D\nu_k=\frac{d\nu_k}{4\pi k}e^{-\frac{i}{4\pi k}\nu_k^2},
\end{align}
while the measures from the one-loop contributions of the vector multiplets and the hypermultiplets are
\begin{align}
V_\text{O}&=\prod_{i<j}^{N_1}\Bigl(2\sinh\frac{\mu_i-\mu_j}{2k}\Bigr)^2
\Bigl(2\sinh\frac{\mu_i+\mu_j}{2k}\Bigr)^2
\prod_{i=1}^{N_1}\Bigl(2\sinh\frac{\mu_i}{2k}\Bigr)^2,\nonumber\\
V_\text{Sp}&=\prod_{k<l}^{N_2}\Bigl(2\sinh\frac{\nu_k-\nu_l}{2k}\Bigr)^2
\Bigl(2\sinh\frac{\nu_k+\nu_l}{2k}\Bigr)^2
\prod_{k=1}^{N_2}\Bigl(2\sinh\frac{\nu_k}{k}\Bigr)^2,\nonumber\\
H&=\prod_{i=1}^{N_1}\prod_{k=1}^{N_2}
\Bigl(2\cosh\frac{\mu_i-\nu_k}{2k}\Bigr)^2
\Bigl(2\cosh\frac{\mu_i+\nu_k}{2k}\Bigr)^2
\prod_{k=1}^{N_2}\Bigl(2\cosh\frac{\nu_k}{2k}\Bigr)^2.
\end{align}
After taking care of the trivial cancellation between $V_\text{Sp}$ and $H$, we find that the partition function is symmetric under the simultaneous exchange of $(N_1,N_2)$ and the sign change of $k$.
Hereafter let us assume $N_1\le N_2$ and $k>0$ without loss of generality and rewrite $Z_k(N_1,N_2)$ as $Z_{k,M}(N)$ by introducing $N=N_1$ and $M=N_2-N_1$.
Otherwise we can simply consider its complex conjugate.

As in the case of the non-equal rank deformation of the ABJM theory \cite{MM}, let us first prepare a determinant formula suitable for the application to the current situation,
\begin{align}
&\det\begin{pmatrix}
\Bigl[\frac{1}
{(z_i+w_k)(1+1/(z_iw_k))}\Bigr]
_{(i,k)\in Z_N\times Z_{N+M}}\\
\Bigl[\frac{w_k^{m-\frac12}-w_k^{-(m-\frac12)}}
{w_k^{\frac12}-w_k^{-\frac12}}
\Bigr]
_{(m,k)\in Z_M\times Z_{N+M}}\\
\end{pmatrix}
=(-1)^{MN+\frac12M(M-1)}
\nonumber\\&\qquad\times
\frac{\prod_{i<j}^{N}(z_i-z_j)(1-1/(z_iz_j))
\prod_{k<l}^{N+M}(w_k-w_l)(1-1/(w_kw_l))}
{\prod_{i=1}^{N}\prod_{k=1}^{N+M}(z_i+w_k)(1+1/(z_iw_k))},
\label{Cauchy}
\end{align}
where $Z_L=\{1,2,\cdots,L\}$ is a set of $L$ elements in this ordering.
This formula can be derived as follows.
We start with the standard Cauchy determinant \cite{MePu,MS1}
\begin{align}
&\det\begin{pmatrix}
\Bigl[\frac{1}{(z_i+w_k)(1+1/(z_iw_k))}\Bigr]
_{(i,k)\in Z_{N+M}\times Z_{N+M}}\\
\end{pmatrix}
\nonumber\\&\qquad
=\frac{\prod_{i<j}^{N+M}(z_i-z_j)(1-1/(z_iz_j))
\prod_{k<l}^{N+M}(w_k-w_l)(1-1/(w_kw_l))}
{\prod_{i=1}^{N+M}\prod_{k=1}^{N+M}(z_i+w_k)(1+1/(z_iw_k))}.
\end{align}
Then, we send $z_{N+1},z_{N+2},\cdots,z_{N+M}$ to infinity one after another using the series expansion in $z$,
\begin{align}
\frac{1}{(z+w)(1+1/(zw))}=\sum_{m=1}^\infty\frac{(-1)^{m-1}}{z^m}
\frac{w^m-w^{-m}}{w-w^{-1}}.
\end{align}
Since in the determinant we can add a multiple of one row to another without changing its value, the leading contribution in the $m$-th row of the lower block is the $z^{-m}$ term, $(w^m-w^{-m})/(w-w^{-1})$.
Note that this coefficient is a Laurent polynomial of $w$.
Again, due to the same property of the row addition, we can keep only the top terms of the polynomials $w^{m-1}+w^{-(m-1)}$ or change the lower terms arbitrarily.
We choose to replace this coefficient by another with half intermediate steps
\begin{align}
\frac{w^m-w^{-m}}{w-w^{-1}}\to
\frac{w^{m-\frac12}-w^{-(m-\frac12)}}
{w^{\frac12}-w^{-\frac12}}.
\end{align}
This proves the determinant formula \eqref{Cauchy}.
Then, after substituting $z_i=e^{\mu_i}$ and $w_k=e^{\nu_k}$ into \eqref{Cauchy}, we can rewrite the measure as the product of two determinants
\begin{align}
\frac{V_\text{O}V_\text{Sp}}{H}
=\det\begin{pmatrix}
\Bigl[\frac{(2\sinh\frac{\mu_i}{2k})(2\sinh\frac{\nu_k}{2k})}
{(2\cosh\frac{\mu_i-\nu_k}{2k})(2\cosh\frac{\mu_i+\nu_k}{2k})}\Bigr]
_{(i,k)\in Z_N\times Z_{N+M}}\\
\Bigl[2\sinh\frac{(m-\frac12)\nu_k}{k}\Bigr]
_{(m,k)\in Z_M\times Z_{N+M}}\\
\end{pmatrix}^2. 
\label{meadet}
\end{align}

As usual, it is useful to introduce the operators $\widehat{q}$ and $\widehat{p}$ satisfying the canonical commutation relation $[\widehat{q},\widehat{p}]=i\hbar$ with the Planck constant identified with $\hbar=2\pi k$. 
In terms of the eigenstates $|\mu\rangle$ of $\widehat{q}$ normalized by $\langle\mu|\nu\rangle=2\pi\delta(\mu-\nu)$, the entries in the upper block of the determinant can be rewritten using the matrix elements 
\begin{align}
\langle\mu_i|\frac{\widehat{\Pi}_-}{2\cosh\frac{\widehat{p}}2}
|\nu_k\rangle
=\frac{1}{2k}\frac{(2\sinh\frac{\mu_i}{2k})(2\sinh\frac{\nu_k}{2k})}
{(2\cosh\frac{\mu_i-\nu_k}{2k})(2\cosh\frac{\mu_i+\nu_k}{2k})}.
\label{cosh}
\end{align}
For the lower entries, we introduce states $\llangle m|$, $|m\rrangle$ defined such that\footnote{
In terms of the suitably normalized zero-momentum eigenstate $|\widetilde 0\rangle$ introduced in \cite{MS1}, this can be expressed as $|m\rrangle=2\sinh\frac{(m-\frac{1}{2})\widehat q}{k}|\widetilde 0\rangle$.
Hence, this state is a linear combination of momentum eigenstates $|\widetilde{p}\rangle$ with imaginary momenta $p=\pm(2m-1)\pi i$.
The subtlety of this state will need a special care later in this section.
} 
\begin{align}
\llangle m|\nu_k\rangle=\langle\nu_k|m\rrangle
=2\sinh\frac{(m-\frac{1}{2})\nu_k}{k}.
\label{doubly}
\end{align}
We can trivialize one of the permutations coming from the determinants in \eqref{meadet} by relabeling the indices of $\nu_k$.
After including the Gaussian factors $e^{\frac{i}{4\pi k}\mu_i^2}$ and $e^{-\frac{i}{4\pi k}\nu_k^2}$, the partition function becomes
\begin{align}
Z_{k,M}(N)
&=\frac{1}{N!}
\int\frac{d^{N}\mu}{(4\pi k)^{N}}\frac{d^{N+M}\nu}{(4\pi k)^{N+M}}
\prod_{i=1}^{N}2k
\langle\mu_i|e^{\frac i{2\hbar}\widehat{q}^2}
\frac{\widehat{\Pi}_-}{2\cosh\frac{\widehat{p}}2}
e^{-\frac i{2\hbar}\widehat{q}^2}|\nu_i\rangle
\prod_{m=1}^M\llangle m|
e^{-\frac i{2\hbar}\widehat{q}^2}|\nu_{N+m}\rangle
\nonumber\\
&\quad\times\det\begin{pmatrix}
\Bigl[2k\langle\nu_k|\frac{\widehat{\Pi}_-}{2\cosh\frac{\widehat{p}}2}
|\mu_j\rangle\Bigr]_{(k,j)\in Z_{N+M}\times Z_N}
\Bigl[\langle\nu_k|n\rrangle\Bigr]_{(k,n)\in Z_{N+M}\times Z_M}
\end{pmatrix}.
\label{Z1}
\end{align}

In the case of equal ranks, it was a standard technique to perform a similarity transformation \cite{HHMO}
\begin{align}
\langle\mu_i|\to\langle\mu_i|e^{\frac i{2\hbar}\widehat{p}^2},\quad
|\mu_i\rangle\to e^{-\frac i{2\hbar}\widehat{p}^2}|\mu_i\rangle,\quad
\langle\nu_k|\to\langle\nu_k|e^{\frac i{2\hbar}\widehat{p}^2},\quad
|\nu_k\rangle\to e^{-\frac i{2\hbar}\widehat{p}^2}|\nu_k\rangle,
\label{similar}
\end{align}
which is allowed because all of these states appear only in
\begin{align}
1=\int\frac{d\mu_i}{2\pi}|\mu_i\rangle\langle\mu_i|
=\int\frac{d\nu_k}{2\pi}|\nu_k\rangle\langle\nu_k|.
\end{align}
Here we follow this similarity transformation and see the effects on each component.
Roughly speaking, in the following we shall see that the matrix elements in the two products in \eqref{Z1} in front of the determinant become delta functions, which enable us to perform the $\nu_k$ integrations.

First, let us consider the determinant part
\begin{align}
\det\begin{pmatrix}
\Bigl[2k\langle\nu_k|e^{\frac i{2\hbar}\widehat{p}^2}
\frac{\widehat{\Pi}_-}{2\cosh\frac{\widehat{p}}2}
e^{-\frac i{2\hbar}\widehat{p}^2}|\mu_j\rangle\Bigr]
_{(k,j)\in Z_{N+M}\times Z_N}&
\Bigl[\langle\nu_k|e^{\frac i{2\hbar}\widehat{p}^2}|n\rrangle\Bigr]
_{(k,n)\in Z_{N+M}\times Z_M}
\end{pmatrix}.
\label{det1}
\end{align}
It is trivial to see the left block of the determinant is unchanged under the similarity transformation, while the right block can be easily computed as
\begin{align}
\langle\nu_k|e^{\frac{i}{2\hbar}\widehat p^2}|n\rrangle
=e^{-\frac{i}{2\hbar}(2\pi(n-\frac{1}{2}))^2}\langle\nu_k|n\rrangle.
\end{align}
After taking care of the extra phase factors, the determinant \eqref{det1} can be written as
\begin{align}
e^{-\frac{\pi i}{12k}M(2M+1)(2M-1)}\det\begin{pmatrix}
\Bigl[2k\langle\nu_k|\frac{\widehat{\Pi}_-}{2\cosh\frac{\widehat{p}}2}
|\mu_j\rangle\Bigr]_{(k,j)\in Z_{N+M}\times Z_N}&
\Bigl[\langle\nu_k|n\rrangle\Bigr]_{(k,n)\in Z_{N+M}\times Z_M}
\end{pmatrix}. 
\label{det2}
\end{align}
Note that this is an odd function of both $\mu_i$ and $\nu_k$ which can be shown by the determinant formula \eqref{Cauchy}.

Next, let us consider the matrix elements in \eqref{Z1} in front of the determinant.
For the first product, after the similarity transformations which changes $(2\cosh\frac{\widehat p}{2})^{-1}$ into $(2\cosh\frac{\widehat q}{2})^{-1}$, we find
\begin{align}
2k\langle\mu_i|e^{\frac i{2\hbar}\widehat{p}^2}
e^{\frac i{2\hbar}\widehat{q}^2}
\frac{\widehat{\Pi}_-}{2\cosh\frac{\widehat{p}}2}
e^{-\frac i{2\hbar}\widehat{q}^2}
e^{-\frac i{2\hbar}\widehat{p}^2}|\nu_i\rangle
=\frac{2\pi k}{2\cosh\frac{\mu_i}{2}}
(\delta(\mu_i-\nu_i)-\delta(\mu_i+\nu_i)),
\label{prodi}
\end{align}
where we have explicitly spelled out the matrix element $\langle\mu_i|\widehat{\Pi}_-|\nu_i\rangle$.
For the second product, we have 
\begin{align}
\llangle m|e^{-\frac{i}{2\hbar}\widehat q^2}
e^{-\frac{i}{2\hbar}\widehat p^2}|\nu_{N+m}\rangle
=\int\frac{d\lambda}{2\pi}
2\sinh\frac{(m-\frac{1}{2})\lambda}{k}
e^{-\frac{i}{2\hbar}\lambda^2}\frac{1}{\sqrt{ik}}
e^{\frac{i}{2\hbar}(\lambda-\nu_{N+m})^2}.
\label{subtle}
\end{align}
There is a subtlety on the definition of this integral which will be clarified at the end of this section.
For the moment, we perform the Gaussian integral formally
\begin{align}
&\llangle m|e^{-\frac{i}{2\hbar}\widehat q^2}
e^{-\frac{i}{2\hbar}\widehat p^2}|\nu_{N+m}\rangle
=\frac{2\pi k}{\sqrt{ik}}
e^{-\frac i{2\hbar}(2\pi(m-\frac12))^2}
\nonumber\\&\quad\times
(\delta(\nu_{N+m}+(2m-1)\pi i)-\delta(\nu_{N+m}-(2m-1)\pi i)).
\label{prodm}
\end{align}
As a result, all the $\nu_k$ integrations can be done explicitly due to the delta functions in \eqref{prodi} and \eqref{prodm}.
There are further simplifications.
Since the remaining determinant \eqref{det2} in the integrand is an odd function of $\nu_k$, we can simply replace the matrix elements discussed above as
\begin{align}
&2k\langle\mu_i|e^{\frac i{2\hbar}\widehat{p}^2}
e^{\frac i{2\hbar}\hat{q}^2}
\frac{\widehat{\Pi}_-}{2\cosh\frac{\widehat{p}}2}
e^{-\frac i{2\hbar}\widehat{q}^2}
e^{-\frac i{2\hbar}\widehat{p}^2}|\nu_i\rangle
\to\frac{4\pi k}{2\cosh\frac{\mu_i}{2}}\delta(\mu_i-\nu_i),\nonumber\\
&\llangle m|e^{-\frac{i}{2\hbar}\widehat q^2}
e^{-\frac{i}{2\hbar}\widehat p^2}|\nu_{N+m}\rangle
\to\frac{4\pi k}{\sqrt{ik}}
e^{-\frac{i}{2\hbar}(2\pi(m-\frac{1}{2}))^2}
\delta\bigl(\nu_{N+m}+(2m-1)\pi i\bigr).
\end{align}

After substituting these replacements and taking care of the extra phase factors, the partition function is given by
\begin{align}
Z_{k,M}(N)&=e^{-\frac{\pi i}{6k}M(2M+1)(2M-1)}(ik)^{-\frac M2}
\frac{1}{N!}
\int\frac{d^N\mu}{(4\pi k)^N}
\prod_{i=1}^N\frac{1}{2\cosh\frac{\mu_i}{2}}
\nonumber\\
&\qquad\times\det\begin{pmatrix}
\Bigl[2k\langle\mu_i|
\frac{\widehat{\Pi}_-}{2\cosh\frac{\widehat{p}}2}|\mu_j\rangle\Bigr]
_{(i,j)\in Z_N\times Z_N}&
\Bigl[\langle\mu_i|n\rrangle\Bigr]
_{(i,n)\in Z_N\times Z_M}\\
\Bigl[2k\langle\rho_m|
\frac{\widehat{\Pi}_-}{2\cosh\frac{\widehat{p}}2}|\mu_j\rangle\Bigr]
_{(m,j)\in Z_M\times Z_N}&
\Bigl[\langle\rho_m|n\rrangle\Bigr]
_{(m,n)\in Z_M\times Z_M}
\end{pmatrix},
\label{Zdet}
\end{align}
where $\rho_m=-(2m-1)\pi i$.
Using again the Cauchy determinant formula \eqref{Cauchy} for the determinant factor in \eqref{Zdet}, finally we find that the partition function is given by 
\begin{align}
\frac{(-1)^{MN}Z_{k,M}(N)}{Z_{k,M}(0)}
&=\frac{1}{N!}\int\frac{d^N\mu}{(4\pi k)^N}
\prod_{i=1}^N\frac{(2\sinh\frac{\mu_i}{2k})^2V(\mu_i)}
{4\cosh\frac{\mu_i}k}
\prod_{i<j}^N
\left(\tanh\frac{\mu_i-\mu_j}{2k}\tanh\frac{\mu_i+\mu_j}{2k}\right)^2,
\label{ZN/Z0}
\end{align}
where we have defined $V(\mu)$ as 
\begin{align}
V(\mu)=\frac1{2\cosh\frac{\mu}2}\prod_{m=1}^M
\tanh\frac{\mu-\rho_m}{2k}\tanh\frac{\mu+\rho_m}{2k},
\label{V}
\end{align}
and the normalization factor as
\begin{align}
Z_{k,M}(0)&=(-1)^{\frac{1}{2}M(M-1)}
e^{-\frac{\pi i}{6k}M(2M+1)(2M-1)}
\left({i}{k}\right)^{-\frac M2}\nonumber\\
&\quad\times\prod_{m=1}^M2\sinh\frac{\rho_m}{2k}
\prod_{m<n}^M4\sinh\frac{\rho_m-\rho_n}{2k}\sinh\frac{\rho_m+\rho_n}{2k}.
\end{align}

The expression \eqref{ZN/Z0} can be interpreted as the partition function of a Fermi gas system
\begin{align}
\frac{(-1)^{MN}Z_{k,M}(N)}{Z_{k,M}(0)}
=\frac{1}{N!}\sum_{\sigma\in S_N}(-1)^\sigma
\int\frac{d^N\mu}{(2\pi)^N}
\prod_{i=1}^N\langle\mu_i|\widehat\rho|\mu_{\sigma(i)}\rangle,
\end{align}
with the density matrix
\begin{align}
\widehat\rho=\sqrt{V(\widehat q)}
\frac{\widehat\Pi_-}{2\cosh\frac{\widehat p}{2}}
\sqrt{V(\widehat q)}.
\end{align}
If we rewrite the function $V(\mu)$ \eqref{V} as
\begin{align}
V(\mu)=\frac{1}{2\cosh\frac{\mu}{2}}
\prod_{s=-(M-\frac{1}{2})}^{M-\frac{1}{2}}
\tanh\frac{\mu+2\pi is}{2k}.
\end{align}
and compare it with the result for U$(N_1|N_2)$, we easily find that this is nothing but (2.21) in \cite{PTEP} with $M$ replaced by $2M$.

Let us now return to the subtlety in \eqref{subtle}.
One way to regularize the integral is to insert $e^{-i\epsilon\lambda^2}$ into \eqref{subtle} with an infinitesimal parameter $\epsilon>0$ and rotate the integration contour clockwise.
Then, the integration becomes
\begin{align}
\llangle m|e^{-\frac{i}{2\hbar}\widehat q^2}
e^{-\frac{i}{2\hbar}\widehat p^2}|\nu_{N+m}\rangle
=\frac{1}{\sqrt{ik}}e^{\frac{i}{2\hbar}\nu_{N+m}^2}
\biggl[\Delta_\epsilon\Bigl(\nu_{N+m},\frac{2m-1}{2k}\Bigr)
-\Delta_\epsilon\Bigl(\nu_{N+m},-\frac{2m-1}{2k}\Bigr)\biggr],
\end{align}
where $\Delta_\epsilon(\nu_{N+m},\alpha)$ is given by
\begin{align}
\Delta_\epsilon(\nu_{N+m},\alpha)
=\int\frac{d\lambda}{2\pi}e^{\alpha\lambda}
e^{-\frac{i}{\hbar}\lambda\nu_{N+m}}e^{-i\epsilon\lambda^2},
\end{align}
which is vanishing in the limit $\epsilon\to 0$ for
\begin{align}
(\re\nu_{N+m})\Bigl(\alpha+\frac{\im\nu_{N+m}}{2\pi k}\Bigr)>0.
\label{vanishcond}
\end{align}
In \eqref{prodm} we have formally rotated $\nu_{N+m}$ counterclockwise to a pure imaginary variable as well and found the integration reduces to a sum of delta functions in the limit $\epsilon\to 0$.
Of course, such a manipulation is allowed only if the integration contour of $\nu_{N+m}$ does not pick up any finite residues in the rotation.
Possible residues might come from poles of the matrix element $2k\langle\nu_{N+m}|\frac{\widehat{\Pi}_-}{2\cosh\frac{\widehat{p}}2}|\mu_j\rangle$ in the determinant in \eqref{Z1}, which are located at $\nu_{N+m}=\pm\mu_j+lk\pi i$ with integral $l$, or more concisely $|\im(\nu_{N+m})|\ge k\pi$, as can be seen from the expression \eqref{cosh}.
On the other hand, our computation \eqref{vanishcond} for the regularized expression shows that the residues in the region $\re(\nu_{N+m})>0,\im(\nu_{N+m})>(2m-1)\pi$ and $\re(\nu_{N+m})<0,\im(\nu_{N+m})<-(2m-1)\pi$ are accompanied by a vanishing factor in the limit $\epsilon\to 0$.
Since the index $m$ runs over $m=1,2,\cdots,M$ and the consistency of the ${\rm OSp}(2N+1|2(N+M))$ theory requires $2M\le k$, only poles in the region $|\im(\nu_{N+m})|<k\pi$ are relevant.
Therefore, we are allowed to use the formal expression \eqref{prodm} in the proof.

\section{Exact functional relation and topological invariants}\label{functional}

In the previous section, we have established the relation between the density matrix for the orthosymplectic OSp$(2N+1|2(N+M))$ (or OSp$(2(N+M)+1|2N)$) matrix model and that for the unitary U$(N|N+2M)$ matrix model with the projection to the odd chirality.
Here we shall proceed to studying the simplest $M=0$ case \cite{Ho2}, the OSp$(2N+1|2N)$ grand potential, which is equivalent to the grand potential $J_{-,k}(\mu)$ constructed from the density matrix for the original ABJM U$(N|N)$ matrix model with the odd projection.
Although the chiral projection of the density matrix was introduced early in \cite{HMO1} and the importance was already stressed in \cite{MePu,MS1}, there has not been a strong motivation to study them carefully\footnote{
Very recently, we are informed by Kazumi Okuyama that the grand potentials of general U$(N_1|N_2)$ theories with the chiral projections are studied \cite{Ok2} in the expectation of its physical relevance.
This section has some overlaps with it.
}
until we know that it appears directly in the orthosymplectic matrix model \cite{Ho2}.
In this section, we shall study the non-perturbative effects of $J_{-,k}(\mu)$ carefully.
We point out a functional relation between the grand potentials with the chiral projections $J_{\pm,k}(\mu)$, from which the membrane instantons due to the chiral projections are determined.
Then, we further turn to the study of the worldsheet instantons in $J_{-,k}(\mu)$.

We first define the grand potentials constructed from the density matrices with the chiral projections
\begin{align}
\sum_{n=-\infty}^\infty e^{J_{\pm,k}(\mu+2\pi in)}=\det(1+e^\mu\rho_\pm).
\end{align}
The perturbative part of each grand potential is given by a cubic polynomial
\begin{align}
J^\text{pert}_{\pm,k}(\mu)
=\frac{C_{\pm,k}}{3}\mu^3+B_{\pm,k}\mu+A_{\pm,k},
\end{align}
with the coefficients related to those of the ABJM theory by
\begin{align}
C_{\pm,k}=\frac{C^\text{ABJM}_{k}}{2},\quad
B_{\pm,k}=\frac{B^\text{ABJM}_{k}\pm1/2}{2},\quad
A_{\pm,k}=\frac{A^\text{ABJM}_{k}\mp\log{2}}{2},
\end{align}
which results in the Airy function as in the full case \cite{FHM}.
Our observation is that the non-perturbative part of the difference between the even and odd grand potentials $J_{\pm,k}(\mu)$ looks quite simple for integral $k$.
After extracting the perturbative terms by
\begin{align}
J_{+,k}(\mu)-J_{-,k}(\mu)
=\frac{\mu}{2}-\log 2+\Delta_k(\mu),
\end{align}
we find that the non-perturbative terms of the difference $\Delta_k(\mu)=J^\text{np}_{+,k}(\mu)-J^\text{np}_{-,k}(\mu)$ satisfy
\begin{align}
&\Delta_{k\equiv 1,7\,\text{mod}\,8}(\mu)
=-\Delta_{k\equiv 3,5\,\text{mod}\,8}(\mu)
=\frac{1}{4}\log\frac{1+2\sqrt{2}e^{-\mu}+4e^{-2\mu}}
{1-2\sqrt{2}e^{-\mu}+4e^{-2\mu}},\nonumber\\
&\Delta_{k\equiv 0\,\text{mod}\,8}(\mu)=\frac{1}{2}\log(1+4e^{-\mu}),
\quad\Delta_{k\equiv 4\,\text{mod}\,8}(\mu)
=\frac{1}{2}\log(1-4e^{-\mu}),\nonumber\\
&\Delta_{k\equiv 2,6\,\text{mod}\,8}(\mu)
=\frac{1}{4}\log(1+16e^{-2\mu}),
\label{npdiff}
\end{align}
from the numerical fitting.
For the reader's convenience, we present in the appendix the exact values of the partition functions and the grand potentials found from the numerical fitting.\footnote{
These exact values are well-known to several experts.
For example, the values for $k=1$ appear in \cite{HMO1} and the values for $k=2,3,4,6$ are the basic ingredients used to compute the values without projections in \cite{HMO2}.
The non-perturbative large $\mu$ expansion of the grand potential should also be known to experts.
For example, some functional relations using them appear in \cite{GHM2}.
The reason that we collect these results here is to justify our functional relation \eqref{npdiff} and to identify the diagonal Gopakumar-Vafa invariants in table \ref{GV}.
}
Note that the expression in \eqref{npdiff} is reminiscent of the odd-power terms of $e^{-\mu}$ in the orthosymplectic OSp$(2N|2N)$ matrix model.
See (2.45) in \cite{MS1}.

In the above, we have seen that the membrane instanton part is corrected for the orthosymplectic matrix model $J_{-,k}(\mu)$.
It is natural to expect that the worldsheet instanton part should be corrected as well if we believe that the total function should have a certain modular invariance connecting the membrane and worldsheet instanton parts.
Since it seems that the membrane instantons do not contain new singularities, we expect that only the worldsheet instantons with genus greater than zero are corrected.
To study the worldsheet instantons carefully, next let us turn to the sum of two grand potentials $J_{\pm,k}(\mu)$, since the difference seems to encode only the membrane instantons.
We first define the non-perturbative effects of the sum $\Sigma_k(\mu_\text{eff})$ as
\begin{align}
J_{+,k}(\mu)+J_{-,k}(\mu)
=\frac{C^\text{ABJM}_k}{3}\mu_\text{eff}^3
+B^\text{ABJM}_k\mu_\text{eff}+A^\text{ABJM}_k+\Sigma_k(\mu_\text{eff}),
\end{align}
where the right-hand side is expressed in terms of the effective chemical potential $\mu_\text{eff}$ given in \cite{HMO3}.
Then, we can rewrite the results in appendix \ref{GP} as in table \ref{sum}.
\begin{table}[!ht]
\begin{align*}
&\Sigma_1(\mu)=\biggl[\frac{8\mu^2+4\mu+1}{4\pi^2}
-\frac{3}{8}\biggr]e^{-4\mu}
+\biggl[-\frac{9(32\mu^2+8\mu+1)}{32\pi^2}+\frac{67}{16}\biggr]e^{-8\mu}
\\&\quad
+\biggl[\frac{41(72\mu^2+12\mu+1)}{54\pi^2}
-\frac{133}{4}\biggr]e^{-12\mu}+{\cal O}(e^{-16\mu}),\\
&\Sigma_2(\mu)=\biggl[\frac{2\mu^2+2\mu+1}{\pi^2}
-\frac{1}{2}\biggr]e^{-2\mu}
+\biggl[-\frac{9(8\mu^2+4\mu+1)}{8\pi^2}+\frac{17}{4}\biggr]e^{-4\mu}
\\&\quad
+\biggl[\frac{82(18\mu^2+6\mu+1)}{27\pi^2}-\frac{101}{3}\biggr]e^{-6\mu}
+\biggl[-\frac{777(32\mu^2+8\mu+1)}{64\pi^2}+\frac{2273}{8}\biggr]
e^{-8\mu}
+{\cal O}(e^{-10\mu}),\\
&\Sigma_3(\mu)=\frac{4}{3}e^{-\frac{4}{3}\mu}
+\biggl[\frac{8\mu^2+4\mu+1}{12\pi^2}-\frac{145}{72}\biggr]e^{-4\mu}
-2e^{-\frac{16}{3}\mu}
+{\cal O}(e^{-\frac{20}{3}\mu}),\\
&\Sigma_4(\mu)=e^{-\mu}
+\biggl[-\frac{2\mu^2+2\mu+1}{2\pi^2}+\frac{5}{2}\biggr]e^{-2\mu}
+\frac{10}{3}e^{-3\mu}
+\biggl[-\frac{9(8\mu^2+4\mu+1)}{16\pi^2}+\frac{49}{4}\biggr]e^{-4\mu}
\\&\quad
+{\cal O}(e^{-5\mu}),\\
&\Sigma_5(\mu)=\frac{2(5-\sqrt{5})}{5}e^{-\frac{4}{5}\mu}
-\frac{5-\sqrt{5}}{5}e^{-\frac{8}{5}\mu}
+\frac{2(5+7\sqrt{5})}{15}e^{-\frac{12}{5}\mu}
+\frac{15-13\sqrt{5}}{10}e^{-\frac{16}{5}\mu}
+{\cal O}(e^{-4\mu}),\\
&\Sigma_6(\mu)=\frac{4}{3}e^{-\frac{2}{3}\mu}
+\biggl[\frac{2\mu^2+2\mu+1}{3\pi^2}-\frac{43}{18}\biggr]e^{-2\mu}
-2e^{-\frac{8}{3}\mu}+{\cal O}(e^{-\frac{10}{3}\mu}),\\
&\Sigma_8(\mu)=2e^{-\frac{1}{2}\mu}-\frac{1}{2}e^{-\mu}
-\frac{4}{3}e^{-\frac{3}{2}\mu}
+\biggl[-\frac{2\mu^2+2\mu+1}{4\pi^2}+\frac{23}{4}\biggr]e^{-2\mu}
+{\cal O}(e^{-\frac{5}{2}\mu}),\\
&\Sigma_{12}(\mu)=4e^{-\frac{1}{3}\mu}
-\frac{8}{3}e^{-\frac{2}{3}\mu}
+\frac{1}{3}e^{-\mu}
+6e^{-\frac{4}{3}\mu}
+{\cal O}(e^{-\frac{5}{3}\mu}).
\end{align*}
\caption{Non-perturbative effects of the sum $\Sigma_{k}(\mu)$ of grand potentials constructed for two chirally projected density matrices.}
\label{sum}
\end{table}

Using the expression of $\Sigma_k(\mu)$ in table \ref{sum}, we find that the coefficients $d_m(k)$ of the worldsheet instantons $e^{-\frac{4m\mu_\text{eff}}{k}}$ for $J^\text{np}_{-,k}(\mu)=(\Sigma_k(\mu_\text{eff})-\Delta_k(\mu))/2$ fit well with the Gopakumar-Vafa formula
\begin{align}
d_m(k)=\frac{(-1)^m}{m}\sum_{g=0}^\infty\sum_{d|m}n^g_d\,
d\Bigl(2\sin\frac{2\pi m}{dk}\Bigr)^{2g-2}.
\end{align}
From the comparison, we can read off the diagonal Gopakumar-Vafa invariants $n^g_d$ directly, which are shown in table \ref{GV}.
It is interesting to note that these invariants are all integers, which is not guaranteed from the beginning.
Here we have listed the invariants for the ABJM theory as well for convenience.
We have found that, as we expected, twice of the invariants for $J_{-,k}(\mu)$ match exactly with those for the ABJM theory for genus zero.
\begin{table}[!ht]
\begin{center}
\begin{tabular}{|c||c|c|c|c|}
\hline
$d$&1&2&3&4\\
\hline\hline
$n_0^d$&$-2$&$-2$&$-6$&$-24$\\
\hline
$n_1^d$& $0$& $1$& $8$& $73$\\
\hline
$n_2^d$& $0$& $0$&$-2$&$-76$\\
\hline
$n_3^d$& $0$& $0$& $0$& $39$\\
\hline
$n_4^d$& $0$& $0$& $0$&$-10$\\
\hline
$n_5^d$& $0$& $0$& $0$&  $1$\\
\hline
$n_6^d$& $0$& $0$& $0$&  $0$\\
\hline
\end{tabular}\qquad
\begin{tabular}{|c||c|c|c|c|}
\hline
$d$&1&2&3&4\\
\hline\hline
$n_0^d$&$-4$&$-4$&$-12$& $-48$\\
\hline
$n_1^d$& $0$& $0$&  $0$&   $9$\\
\hline
$n_2^d$& $0$& $0$&  $0$&   $0$\\
\hline
$n_3^d$& $0$& $0$&  $0$&   $0$\\
\hline
$n_4^d$& $0$& $0$&  $0$&   $0$\\
\hline
$n_5^d$& $0$& $0$&  $0$&   $0$\\
\hline
$n_6^d$& $0$& $0$&  $0$&   $0$\\
\hline
\end{tabular}
\caption{The diagonal Gopakumar-Vafa invariants identified for the chirally projected model $J_{-,k}(\mu)$ (left) and the ABJM matrix model (right).}
\label{GV}
\end{center}
\end{table}

In principle the diagonal Gopakumar-Vafa invariants come from the trivial relation
\begin{align}
\sum_{n=-\infty}^\infty e^{J^\text{ABJM}_k(\mu+2\pi in)}
=\Biggl[\sum_{n_+=-\infty}^\infty e^{J_{+,k}(\mu+2\pi in_+)}\Biggr]
\Biggl[\sum_{n_-=-\infty}^\infty e^{J_{-,k}(\mu+2\pi in_-)}\Biggr],
\label{sumGP}
\end{align}
between two chirally projected grand potentials.
It would be interesting to derive the invariants directly from \eqref{sumGP}.

\section{Conclusion and discussion}

In this paper we have shown that the claim \cite{Ho2} that the density matrix for the OSp$(2N+1|2N)$ matrix model matches with that for the U$(N|N)$ with the odd chiral projection is extended to \eqref{oddprojM}, after the inclusion of the fractional brane.
We have also proceeded to study the grand potentials constructed from the density matrices projected to the even and odd chiralities, where we find a functional relation which determines the new membrane instanton effects.
We have further studied the worldsheet instanton effects and identified the first few diagonal Gopakumar-Vafa invariants.

We have restricted ourselves to the study of the non-equal rank deformation of the OSp$(2N+1|2N)$ density matrix.
It is apparently interesting to see the same non-equal rank deformation of the OSp$(2N|2N)$ density matrix \cite{MS1} and/or the BPS Wilson loop one-point function in these theories along the line of \cite{HHMO,MM}.
It is interesting to find that, as a general rule, the orientifold projection used to construct the orthosymplectic Chern-Simons theories from the unitary one seems to have a relation to the chiral projection of the corresponding density matrix appearing in the Fermi gas formalism of the matrix model.
We would like to see the physical interpretation of this fact more directly.

\section*{Acknowledgements}
We are grateful to Masazumi Honda for explaining his result to us at the YITP workshop ``Strings and Fields'' prior to the publication \cite{Ho2}.
We would also like to thank Yasuyuki Hatsuda, Shinji Hirano, Takuya Matsumoto, Tomoki Nosaka, Kazumi Okuyama, Masaki Shigemori for valuable discussions.
The work of S.M.\ is supported by JSPS Grant-in-Aid for Scientific
Research (C) \# 26400245.
S.M.\ would like to thank Yukawa Institute for Theoretical Physics at Kyoto University for hospitality.

\appendix

\section{Chirally projected density matrices}

\subsection{Exact values of the partition functions}

In this appendix, we record the first few exact values of the ABJM partition functions with the projections to the even and odd chiralities.
They are given respectively in tables \ref{posPF} and \ref{negPF}.

\begin{table}[!ht]
\footnotesize
\begin{align*}
&Z_{+,1}(1)=\frac{1}{4\sqrt{2}},\quad
Z_{+,1}(2)=\frac{-2+\pi}{64\pi},\quad
Z_{+,1}(3)=\frac{-2\sqrt{2}+(8-5\sqrt{2})\pi}{512\pi},
\\
&Z_{+,2}(1)=\frac{2+\pi}{16\pi},\quad
Z_{+,2}(2)=\frac{3(-2+\pi)^2}{512\pi^2},\quad
Z_{+,2}(3)=\frac{168+396\pi-58\pi^2-27\pi^3}{73728\pi^3},
\\
&Z_{+,3}(1)=\frac{2-\sqrt{2}}{8},\quad
Z_{+,3}(2)=\frac{-18+(27-54\sqrt{2}+32\sqrt{3})\pi}{1728\pi},
\\&\quad
Z_{+,3}(3)=\frac{18(14+\sqrt{2})
-(90-135\sqrt{2}+64\sqrt{3}+32\sqrt{6})\pi}{13824\pi},
\\
&Z_{+,4}(1)=\frac{-1+2\sqrt{2}}{32},\quad
Z_{+,4}(2)=\frac{-16+32\sqrt{2}\pi-(7+4\sqrt{2})\pi^2}{2048\pi^2},
\\&\quad
Z_{+,4}(3)=\frac{16-160\sqrt{2}-32\sqrt{2}\pi+5(-7+10\sqrt{2})\pi^2}
{65536\pi^2},
\\
&Z_{+,5}(1)=\frac{-5\sqrt{2}+4\sqrt{5}}{40},\quad
Z_{+,5}(2)=\frac{150+(625-100\sqrt{10}-16(5\sqrt{2}+2\sqrt{10})
\sqrt{5-\sqrt{5}})\pi}{8000},
\\&\quad
Z_{+,5}(3)=\Bigl[-50(3\sqrt{2}+4\sqrt{5})
+(200+125\sqrt{2}-300\sqrt{5}+(64\sqrt{10}+128\sqrt{2})\sqrt{5-\sqrt{5}}
\\&\qquad\qquad\qquad
+16(5-6\sqrt{2}+3\sqrt{5}-2\sqrt{10})\sqrt{5+\sqrt{5}})\pi\Bigr]
/\bigl[64000\pi\bigr],
\\
&Z_{+,6}(1)=\frac{-18+(9+8\sqrt{3})\pi}{432\pi},\quad
Z_{+,6}(2)=\frac{756-12(189+8\sqrt{3})\pi+(949-144\sqrt{3})\pi^2}
{124416\pi^2},
\\&\quad
Z_{+,6}(3)=\frac{-36936+2268(81+8\sqrt{3})\pi+54(4451+720\sqrt{3})\pi^2
-(37503+46792\sqrt{3})\pi^3}{161243136\pi^3},
\\
&Z_{+,8}(1)=\frac{5-4\sqrt{2-\sqrt{2}}}{64},\quad
Z_{+,8}(2)=\frac{-32+64\sqrt{2+\sqrt{2}}\pi
+(17-32\sqrt{2}-8\sqrt{2-\sqrt{2}})\pi^2}{8192\pi^2},
\\&\quad
Z_{+,8}(3)=\frac{32(-5+12\sqrt{2-\sqrt{2}})
+64(-8\sqrt{2}+\sqrt{2+\sqrt{2}})\pi
+(727-160\sqrt{2}-420\sqrt{2-\sqrt{2}})\pi^2}{524288\pi^2},
\\
&Z_{+,12}(1)=\frac{-5-2\sqrt{2}+4\sqrt{6}}{96},\quad
Z_{+,12}(2)=\frac{-432+288(\sqrt{2}+2\sqrt{6})\pi
+(889+564\sqrt{2}-864\sqrt{3}-296\sqrt{6})\pi^2}{165888\pi^2},
\\&\quad
Z_{+,12}(3)=\Bigl[432(15+14\sqrt{2}-28\sqrt{6})
-96(-216+141\sqrt{2}+74\sqrt{6})\pi
\\&\qquad\qquad\qquad
+(-49101-37654\sqrt{2}+10656\sqrt{3}+36908\sqrt{6})\pi^2\Bigr]
/\bigl[47775744\pi^2\bigr]
\end{align*}
\caption{Exact values of the partition function $Z_{+,k}(N)$ of the ABJM theory with the projection to the even chirality.}
\label{posPF}
\end{table}

\begin{table}[!ht]
\footnotesize
\begin{align*}
&Z_{-,1}(1)=\frac{2-\sqrt{2}}{8},\quad
Z_{-,1}(2)=\frac{6+(1-2\sqrt{2})\pi}{64\pi},\quad
Z_{-,1}(3)=\frac{-20-6\sqrt{2}+(2+5\sqrt{2})\pi}{512\pi},
\\
&Z_{-,2}(1)=\frac{-2+\pi}{16\pi},\quad
Z_{-,2}(2)=\frac{12+12\pi-5\pi^2}{512\pi^2},\quad
Z_{-,2}(3)=\frac{-168+396\pi+202\pi^2-99\pi^3}{73728\pi^3},
\\
&Z_{-,3}(1)=\frac{-4+3\sqrt{2}}{24},\quad
Z_{-,3}(2)=\frac{-90+(135-36\sqrt{2}-32\sqrt{3})\pi}{1728\pi},
\\&\quad
Z_{-,3}(3)=\frac{-72-90\sqrt{2}+(180-27\sqrt{2}-32\sqrt{6})\pi}
{13824\pi},
\\
&Z_{-,4}(1)=\frac{3-2\sqrt{2}}{32},\quad
Z_{-,4}(2)=\frac{-16-32\sqrt{2}\pi+(33-12\sqrt{2})\pi^2}{2048\pi^2},
\\&\quad
Z_{-,4}(3)=\frac{-48+160\sqrt{2}-96\sqrt{2}\pi+(209-130\sqrt{2})\pi^2}
{65536\pi^2},
\\
&Z_{-,5}(1)=\frac{2+5\sqrt{2}-4\sqrt{5}}{40},\quad
Z_{-,5}(2)=\frac{350+(-25(-9-2\sqrt{2}+8\sqrt{5}+4\sqrt{10})
+16(5+\sqrt{5})\sqrt{5+2\sqrt{5}})\pi}{8000\pi},
\\&\quad
Z_{-,5}(3)=\Bigl[50(-2+35\sqrt{2}+4\sqrt{5})+(125(42-21\sqrt{2}+12\sqrt{5}-8\sqrt{10})
\\&\qquad\qquad\qquad
+16(-270+25\sqrt{2}+74\sqrt{5}+5\sqrt{10})\sqrt{5+2\sqrt{5}})\pi\Bigr]
/\bigl[320000\pi\bigr],
\\
&Z_{-,6}(1)=\frac{18+(9-8\sqrt{3})\pi}{432\pi},\quad
Z_{-,6}(2)=\frac{756+(2268-96\sqrt{3})\pi+(-995+144\sqrt{3})\pi^2}
{124416\pi^2},
\\&\quad
Z_{-,6}(3)=\frac{36936-2268(-81+8\sqrt{3})\pi+270(-1279+144\sqrt{3})\pi^2
+(-89991+93448\sqrt{3})\pi^3}{161243136\pi^3},
\\
&Z_{-,8}(1)=\frac{-3+4\sqrt{2-\sqrt{2}}}{64},\quad
Z_{-,8}(2)=\frac{-32-64\sqrt{2+\sqrt{2}}\pi
+(129-32\sqrt{2}-56\sqrt{2-\sqrt{2}})\pi^2}{8192\pi^2},
\\&\quad
Z_{-,8}(3)=\frac{96-384\sqrt{2-\sqrt{2}}
+64(-8\sqrt{2}+7\sqrt{2+\sqrt{2}})\pi
+(-1009+96\sqrt{2}+1124\sqrt{2-\sqrt{2}})\pi^2}{524288\pi^2},
\\
&Z_{-,12}(1)=\frac{7+2\sqrt{2}-4\sqrt{6}}{96},\quad
Z_{-,12}(2)=\frac{-432-288(\sqrt{2}+2\sqrt{6})\pi
+(3697-132\sqrt{2}-864\sqrt{3}-568\sqrt{6})\pi^2}{165888\pi^2},
\\&\quad
Z_{-,12}(3)=\Bigl[3024(-3-2\sqrt{2}+4\sqrt{6})
+96(216+33\sqrt{2}-142\sqrt{6})\pi
\\&\qquad\qquad\qquad
+(157863+40678\sqrt{2}-20448\sqrt{3}-72908\sqrt{6})\pi^2\Bigr]
/\bigl[47775744\pi^2\bigr].
\end{align*}
\caption{Exact values of the partition function $Z_{-,k}(N)$ of the ABJM theory with the projection to the odd chirality.}
\label{negPF}
\end{table}

\subsection{Grand Potential}\label{GP}

In this appendix, we shall present the grand potentials of the ABJM matrix model with the projections to the even and odd chiralities.
They are given respectively in tables \ref{posGP} and \ref{negGP}.

\begin{table}[!p]
\footnotesize
\begin{align*}
&J^\text{np}_{+,1}
=\frac{1}{\sqrt{2}}e^{-\mu}
-\frac{4}{3\sqrt{2}}e^{-3\mu}
+\biggl[\frac{2\mu^2+\mu/2+1/8}{\pi^2}\biggr]e^{-4\mu}
-\frac{16}{5\sqrt{2}}e^{-5\mu}+\frac{64}{7\sqrt{2}}e^{-7\mu}
\nonumber\\&\quad
+\biggl[-\frac{13\mu^2+\mu/8+9/64}{\pi^2}+2\biggr]e^{-8\mu}
+\frac{256}{9\sqrt{2}}e^{-9\mu}-\frac{1024}{11\sqrt{2}}e^{-11\mu}
\nonumber\\&\quad
+\biggl[\frac{368\mu^2-76\mu/3+77/36}{3\pi^2}-32\biggr]e^{-12\mu}
-\frac{4096}{13\sqrt{2}}e^{-13\mu}
+\frac{16384}{15\sqrt{2}}e^{-15\mu}+{\cal O}(e^{-16\mu}),\\
&J^\text{np}_{+,2}
=\biggl[\frac{2\mu^2+\mu+1/2}{\pi^2}+2\biggr]e^{-2\mu}
+\biggl[-\frac{13\mu^2+\mu/4+9/16}{\pi^2}-14\biggr]e^{-4\mu}
\nonumber\\&\quad
+\biggl[\frac{368\mu^2-152\mu/3+77/9}{3\pi^2}
+\frac{416}{3}\biggr]e^{-6\mu}
+\biggl[-\frac{2701\mu^2-13949\mu/24+11291/192}{2\pi^2}
-1582\biggr]e^{-8\mu}
\\&\quad
+\biggl[\frac{80912\mu^2-317122\mu/15+285253/150}{5\pi^2}
+\frac{97472}{5}\biggr]e^{-10\mu}
+{\cal O}(e^{-12\mu}),\\
&J^\text{np}_{+,3}
=-\frac{1}{\sqrt{2}}e^{-\mu}
+\frac{2}{3}e^{-\frac{4}{3}\mu}
+\frac{4}{3\sqrt{2}}e^{-3\mu}
+\biggl[\frac{2\mu^2+\mu/2+1/8}{3\pi^2}-\frac{8}{9}\biggr]e^{-4\mu}
+\frac{16}{5\sqrt{2}}e^{-5\mu}
-\frac{17}{9}e^{-\frac{16}{3}\mu}
\\&\quad
+\frac{2}{15}e^{-\frac{20}{3}\mu}
-\frac{64}{7\sqrt{2}}e^{-7\mu}
+\biggl[-\frac{13\mu^2+\mu/8+9/64}{3\pi^2}+\frac{88}{9}\biggr]e^{-8\mu}
-\frac{256}{9\sqrt{2}}e^{-9\mu}
+{\cal O}(e^{-\frac{28}{3}\mu}),\\
&J^\text{np}_{+,4}
=-\frac{1}{2}e^{-\mu}
+\biggl[-\frac{2\mu^2+\mu+1/2}{2\pi^2}-1\biggr]e^{-2\mu}
-\frac{8}{3}e^{-3\mu}
+\biggl[-\frac{13\mu^2+\mu/4+9/16}{2\pi^2}-6\biggr]e^{-4\mu}
\nonumber\\&\quad
-\frac{128}{5}e^{-5\mu}
+\biggl[-\frac{184\mu^2-76\mu/3+77/18}{3\pi^2}
-\frac{160}{3}\biggr]e^{-6\mu}
-\frac{2048}{7}e^{-7\mu}+{\cal O}(e^{-8\mu}),\\
&J^\text{np}_{+,5}
=\frac{5-\sqrt{5}}{5}e^{-\frac{4}{5}\mu}
-\frac{1}{\sqrt{2}}e^{-\mu}
-\frac{5-\sqrt{5}}{10}e^{-\frac{8}{5}\mu}
+\frac{5+7\sqrt{5}}{15}e^{-\frac{12}{5}\mu}
+\frac{4}{3\sqrt{2}}e^{-3\mu}
+\frac{15-13\sqrt{5}}{20}e^{-\frac{16}{5}\mu}\\
&\quad
+\biggl[\frac{2\mu^2+\mu/2+1/8}{5\pi^2}
+\frac{-13+5\sqrt{5}}{5}\biggr]e^{-4\mu}
-\frac{145+131\sqrt{5}}{150}e^{-\frac{24}{5}\mu}
+\frac{16}{5\sqrt{2}}e^{-5\mu}
+{\cal O}(e^{-\frac{28}{5}\mu}),\\
&J^\text{np}_{+,6}
=\frac{2}{3}e^{-\frac{2}{3}\mu}
+\biggl[\frac{2\mu^2+\mu+1/2}{3\pi^2}+\frac{10}{9}\biggr]e^{-2\mu}
-\frac{17}{9}e^{-\frac{8}{3}\mu}
+\frac{2}{15}e^{-\frac{10}{3}\mu}
+\biggl[-\frac{13\mu^2+\mu/4+9/16}{3\pi^2}-\frac{56}{9}\biggr]e^{-4\mu}
\nonumber\\&\quad
+\frac{2776}{189}e^{-\frac{14}{3}\mu}
-\frac{31}{18}e^{-\frac{16}{3}\mu}
+\biggl[\frac{368\mu^2-152\mu/3+77/9}{9\pi^2}
+\frac{1408}{27}\biggr]e^{-6\mu}
-\frac{35938}{243}e^{-\frac{20}{3}\mu}
+\frac{6508}{297}e^{-\frac{22}{3}\mu}
\nonumber\\&\quad
+\biggl[-\frac{2701\mu^2-13949\mu/24+11291/192}{6\pi^2}
-\frac{4648}{9}\biggr]e^{-8\mu}
+{\cal O}(e^{-\frac{26}{3}\mu}),\\
&J^\text{np}_{+,8}
=e^{-\frac{1}{2}\mu}
+\frac{3}{4}e^{-\mu}
-\frac{2}{3}e^{-\frac{3}{2}\mu}
+\biggl[-\frac{2\mu^2+\mu+1/2}{4\pi^2}+\frac{1}{2}\biggr]e^{-2\mu}
+\frac{6}{5}e^{-\frac{5}{2}\mu}
+4e^{-3\mu}
-\frac{20}{7}e^{-\frac{7}{2}\mu}
\\&\quad
+\biggl[-\frac{13\mu^2+\mu/4+9/16}{4\pi^2}+6\biggr]e^{-4\mu}
+\frac{70}{9}e^{-\frac{9}{2}\mu}
+\frac{192}{5}e^{-5\mu}
-\frac{252}{11}e^{-\frac{11}{2}\mu}
\\&\quad
+\biggl[-\frac{92\mu^2-38\mu/3+77/36}{3\pi^2}
+\frac{224}{3}\biggr]e^{-6\mu}
+\frac{924}{13}e^{-\frac{13}{2}\mu}
+\frac{3072}{7}e^{-7\mu}+{\cal O}(e^{-\frac{15}{2}\mu}),\\
&J^\text{np}_{+,12}
=2e^{-\frac{1}{3}\mu}
-\frac{4}{3}e^{-\frac{2}{3}\mu}
-\frac{5}{6}e^{-\mu}
+3e^{-\frac{4}{3}\mu}
-\frac{38}{5}e^{-\frac{5}{3}\mu}
+\biggl[-\frac{2\mu^2+\mu+1/2}{6\pi^2}+\frac{127}{9}\biggr]e^{-2\mu}
-\frac{344}{21}e^{-\frac{7}{3}\mu}
\\&\quad
+\frac{265}{18}e^{-\frac{8}{3}\mu}
-\frac{40}{9}e^{-3\mu}
-\frac{514}{15}e^{-\frac{10}{3}\mu}
+\frac{3196}{33}e^{-\frac{11}{3}\mu}
+\biggl[-\frac{13\mu^2+\mu/4+9/16}{6\pi^2}
-\frac{1552}{9}\biggr]e^{-4\mu}
+{\cal O}(e^{-\frac{13}{3}\mu}).
\end{align*}
\caption{Non-perturbative effects of the grand potential $J^\text{np}_{+,k}(\mu)$ with the projection to the even chirality.}
\label{posGP}
\end{table}

\begin{table}[!p]
\footnotesize
\begin{align*}
&J^\text{np}_{-,1}=-\frac{1}{\sqrt{2}}e^{-\mu}
+\frac{4}{3\sqrt{2}}e^{-3\mu}
+\biggl[\frac{2\mu^2+\mu/2+1/8}{\pi^2}\biggr]e^{-4\mu}
+\frac{16}{5\sqrt{2}}e^{-5\mu}
-\frac{64}{7\sqrt{2}}e^{-7\mu}
\\&\quad
+\biggl[-\frac{13\mu^2+\mu/8+9/64}{\pi^2}+2\biggr]e^{-8\mu}
-\frac{256}{9\sqrt{2}}e^{-9\mu}
+\frac{1024}{11\sqrt{2}}e^{-11\mu}
\\&\quad
+\biggl[\frac{368\mu^2-76\mu/3+77/36}{3\pi^2}-32\biggr]e^{-12\mu}
+\frac{4096}{13\sqrt{2}}e^{-13\mu}
-\frac{16384}{15\sqrt{2}}e^{-15\mu}+{\cal O}(e^{-16\mu}),\\
&J^\text{np}_{-,2}
=\biggl[\frac{2\mu^2+\mu+1/2}{\pi^2}-2\biggr]e^{-2\mu}
+\biggl[-\frac{13\mu^2+\mu/4+9/16}{\pi^2}+18\biggr]e^{-4\mu}
\nonumber\\&\quad
+\biggl[\frac{368\mu^2-152\mu/3+77/9}{3\pi^2}
-\frac{608}{3}\biggr]e^{-6\mu}
+\biggl[-\frac{2701\mu^2-13949\mu/24+11291/192}{2\pi^2}
+2514\biggr]e^{-8\mu}
\\&\quad
+\biggl[\frac{80912\mu^2-317122\mu/15+285253/150}{5\pi^2}
-\frac{164672}{5}\biggr]e^{-10\mu}
+{\cal O}(e^{-12\mu}),\\
&J^\text{np}_{-,3}=\frac{1}{\sqrt{2}}e^{-\mu}
+\frac{2}{3}e^{-\frac{4}{3}\mu}
-\frac{4}{3\sqrt{2}}e^{-3\mu}
+\biggl[\frac{2\mu^2+\mu/2+1/8}{3\pi^2}-\frac{8}{9}\biggr]e^{-4\mu}
-\frac{16}{5\sqrt{2}}e^{-5\mu}
-\frac{17}{9}e^{-\frac{16}{3}\mu}
\\&\quad
+\frac{2}{15}e^{-\frac{20}{3}\mu}
+\frac{64}{7\sqrt{2}}e^{-7\mu}
+\biggl[-\frac{13\mu^2+\mu/8+9/64}{3\pi^2}+\frac{88}{9}\biggr]e^{-8\mu}
+\frac{256}{9\sqrt{2}}e^{-9\mu}+{\cal O}(e^{-\frac{28}{3}\mu}),\\
&J^\text{np}_{-,4}=\frac{3}{2}e^{-\mu}
+\biggl[-\frac{2\mu^2+\mu+1/2}{2\pi^2}+3\biggr]e^{-2\mu}
+8e^{-3\mu}
+\biggl[-\frac{13\mu^2+\mu/4+9/16}{2\pi^2}+26\biggr]e^{-4\mu}
\\&\quad
+\frac{384}{5}e^{-5\mu}
+\biggl[-\frac{184\mu^2-76\mu/3+77/18}{3\pi^2}
+\frac{864}{3}\biggr]e^{-6\mu}
+\frac{6144}{7}e^{-7\mu}+{\cal O}(e^{-8\mu}),\\
&J^\text{np}_{-,5}
=\frac{5-\sqrt{5}}{5}e^{-\frac{4}{5}\mu}
+\frac{1}{\sqrt{2}}e^{-\mu}
-\frac{5-\sqrt{5}}{10}e^{-\frac{8}{5}\mu}
+\frac{5+7\sqrt{5}}{15}e^{-\frac{12}{5}\mu}
-\frac{4}{3\sqrt{2}}e^{-3\mu}
+\frac{15-13\sqrt{5}}{20}e^{-\frac{16}{5}\mu}
\\&\quad
+\biggl[\frac{2\mu^2+\mu/2+1/8}{5\pi^2}
+\frac{-13+5\sqrt{5}}{5}\biggr]e^{-4\mu}
-\frac{145+131\sqrt{5}}{150}e^{-\frac{24}{5}\mu}
-\frac{16}{5\sqrt{2}}e^{-5\mu}+{\cal O}(e^{-\frac{28}{5}\mu}),\\
&J^\text{np}_{-,6}=\frac{2}{3}e^{-\frac{2}{3}\mu}
+\biggl[\frac{2\mu^2+\mu+1/2}{3\pi^2}-\frac{26}{9}\biggr]e^{-2\mu}
-\frac{17}{9}e^{-\frac{8}{3}\mu}
+\frac{2}{15}e^{-\frac{10}{3}\mu}
+\biggl[-\frac{13\mu^2+\mu/4+9/16}{3\pi^2}+\frac{232}{9}\biggr]e^{-4\mu}
\\&\quad
+\frac{2776}{189}e^{-\frac{14}{3}\mu}
-\frac{31}{18}e^{-\frac{16}{3}\mu}
+\biggl[\frac{368\mu^2-152\mu/3+77/9}{9\pi^2}
-\frac{7808}{27}\biggr]e^{-6\mu}
-\frac{35938}{243}e^{-\frac{20}{3}\mu}
+\frac{6508}{297}e^{-\frac{22}{3}\mu}
\\&\quad
+\biggl[-\frac{2701\mu^2-13949\mu/24+11291/192}{6\pi^2}
+\frac{32216}{9}\biggr]e^{-8\mu}
+{\cal O}(e^{-\frac{26}{3}\mu}),\\
&J^\text{np}_{-,8}
=e^{-\frac{1}{2}\mu}-\frac{5}{4}e^{-\mu}-\frac{2}{3}e^{-\frac{3}{2}\mu}
+\biggl[-\frac{2\mu^2+\mu+1/2}{4\pi^2}+\frac{9}{2}\biggr]e^{-2\mu}
+\frac{6}{5}e^{-\frac{5}{2}\mu}-\frac{20}{3}e^{-3\mu}
-\frac{20}{7}e^{-\frac{7}{2}\mu}
\\&\quad
+\biggl[-\frac{13\mu^2+\mu/4+9/16}{4\pi^2}+38\biggr]e^{-4\mu}
+\frac{70}{9}e^{-\frac{9}{2}\mu}-64e^{-5\mu}
-\frac{252}{11}e^{-\frac{11}{2}\mu}
\\&\quad
+\biggl[-\frac{92\mu^2-38\mu/3+77/36}{3\pi^2}+416\biggr]e^{-6\mu}+\frac{924}{13}e^{-\frac{13}{2}\mu}
-\frac{5120}{7}e^{-7\mu}+{\cal O}(e^{-\frac{15}{2}\mu}),\\
&J^\text{np}_{-,12}=2e^{-\frac{1}{3}\mu}
-\frac{4}{3}e^{-\frac{2}{3}\mu}
+\frac{7}{6}e^{-\mu}
+3e^{-\frac{4}{3}\mu}
-\frac{38}{5}e^{-\frac{5}{3}\mu}
+\biggl[-\frac{2\mu^2+\mu+1/2}{6\pi^2}+\frac{163}{9}\biggr]e^{-2\mu}
-\frac{344}{21}e^{-\frac{7}{3}\mu}
\\&\quad
+\frac{265}{18}e^{-\frac{8}{3}\mu}
+\frac{56}{9}e^{-3\mu}
-\frac{514}{15}e^{-\frac{10}{3}\mu}
+\frac{3196}{33}e^{-\frac{11}{3}\mu}
+\biggl[-\frac{13\mu^2+\mu/4+9/16}{6\pi^2}-\frac{1264}{9}\biggr]e^{-4\mu}
+{\cal O}(e^{-\frac{13}{3}\mu}).
\end{align*}
\caption{Non-perturbative effects of the grand potential $J^\text{np}_{-,k}(\mu)$ with the projection to the odd chirality.}
\label{negGP}
\end{table}

\end{document}